\documentclass[aps,prb,reprint]{revtex4-1}
\usepackage{graphicx}
\usepackage{dcolumn}
\usepackage{bm}
\usepackage{amssymb}
\usepackage{amsmath}
\usepackage{epsf}
\usepackage{color}
\draft

\begin{document}
\preprint{APS/123-QED}

\title{Electric field induced Mott-insulator to metal transition and memristive behaviour in epitaxial V$_2$O$_3$ thin film}

\author{Binoy Krishna De$^1$, V. G. Sathe$^1$, S. B. Roy$^{1,2}$}
\email{sindhunilbroy@gmail.com}
 \affiliation{%
 $^1$UGC DAE Consortium for Scientific Research\\
 Indore-452001, India \\
$^2$Ramakrishna Mission Vivekananda Education and Research Institute\\Belur Math, Howrah-711202, India}%

\date{\today}
\begin{abstract}
We report an isothermal electric field-induced first order phase transition from Mott-insulator to the metallic state in the epitaxial thin film of V$_2$O$_3$ in the temperature regime below its Mott transition temperature $\approx$ 180 K. This isothermal electric field induced transition is accompanied by interesting electro-thermal history effects, which depend on the measurement paths followed in the electric field - temperature phase space. These interesting properties result in tuneable resistive switching and distinct memristive behaviour in V$_2$O$_3$. A generalized framework of disorder-influenced first order phase transition in combination with a resistor network model has been used to explain the observed experimental features. These findings promise possibilities for Mott insulators to be used as highly energy-efficient switches in novel technologies like neuromorphic computing.
\end{abstract}
                             
\maketitle
\section{Introduction}
The archetypal Mott-insulator V$_2$O$_3$ undergoes a transition from a paramagnetic metallic state to a low-temperature antiferromagnetic insulating state around a temperature $T_N \approx$ 160 K \cite{imada}. This temperature induced first-order metal-insulator transition in V$_2$O$_3$ was first observed by Foex (see ref.\cite{yethi} and reference within) in 1946 that is even before Nevill Mott \cite{mott1} advanced his theory of strongly correlated electron systems. A phase diagram for a long-range Coulomb interaction-driven first-order transition between a degenerate electron gas and a localized antiferromagnetic insulator at low temperatures was proposed in a subsequent study by Mott \cite{mott2,mott3}. This first-order phase transition arises out of a thermally perturbed balance of entropy, kinetic energy, and Coulomb repulsion among conduction electrons \cite{imada,rozen}, and leads to an abrupt electron localization \cite{stewart}. From the 1960s pure and doped V$_2$O$_3$ have been extensively studied to understand various aspects of Mott metal-insulator transition \cite{mcw1,mcw2,rice,mcw3}. Yethiraj \cite{yethi} has recorded a concise review of the experimental works on pure and doped-V$_2$O$_3$ leading up to 1990.   

Subsequent experimental studies have revealed newer aspects associated with the Mott metal-insulator transition in V$_2$O$_3$, which suggested a more subtle phase diagram where real-space phase inhomogeneities play an important role \cite{stewart,lupi,frand}. McLeod et al \cite{macl} studied the real-space evolution of the insulator-metal transition in a thin-film V$_2$O$_3$ sample by high spatial-resolution imaging with near-field infrared microscopy, and observed spontaneously nanotextured coexistence of correlated metal and Mott insulator phases in the temperature regime ($\approx$ 160-180 K) \cite{macl}. The newly discovered properties of Mott transition in V$_2$O$_3$  and their sensitivity to external stimuli make it promising for applications, such as memory, selectors for ReRAM, optical switches, and emulating brain functionalities (see ref. \cite{kolch} and references therein). The easiest and most practical way of inducing the insulator-to-metal transition in electronic devices is by applying electrical current or voltage. The possibility of an electric field-driven gap collapse in a Mott insulator can lead to a much larger carrier density than in the usual band semiconductors and potentially overcome many limitations of conventional electronic devices \cite{mazza}. Thus the electric field induced insulator to metal transition in various Mott insulators including V$_2$O$_3$ has become a subject of intensive study during the last two decades \cite{mazza,oka,sugi,valm,valle}.

Here we report interesting electro-thermal irreversibility associated with the electric-field induced first order Mott-insulator to correlated metal transition (IMT) in an epitaxial thin film of V$_2$O$_3$. Beyond a critical electric field, the sample shows a tunable memory effect namely memristive behaviour, which depends on the temperature and electric field cycle. To the best of our knowledge, this kind of definitive memristive effect with several minor hysteresis loops in voltage (V) - current (I) curves has not been reported so far in V$_2$O$_3$ and for that matter in any Mott insulators or band semiconductors. This memory effect is robust, and thus it further enhances the possibility of the usage of V$_2$O$_3$ as a two-port tunable electrical switching device with low power consumption. A framework of a disordered influenced first order phase transition \cite{imrywortis} has been used earlier to explain magnetic field-induced functional properties in several classes of materials \cite{sbroy1}. We use this framework along with a resistor network model to explain the observed electric field-induced functional behaviour in the present thin film of V$_2$O$_3$ \cite{stol1,stol2}.  A preliminary version of this work has been reported earlier \cite{condmat}.

\section{Experimental}
The V$_2$O$_3$ thin film was deposited on (222) Al$_2$O$_3$ substrate using the Pulse Laser deposition (PLD) method. A sintered V$_2$O$_5$ pellet was used as a target and 248 nm KrF pulse Excimer laser of pulse energy 2J/cm$^2$ and repetition rate 3 Hz was used for deposition. Sapphire (Al$_2$O$_3$) substrate was kept at a distance of 5 cm from the target. During deposition, the substrate temperature was 650$^0$C and the deposition chamber was evacuated to 10$^{-6}$ mbar.  After deposition, the sample was cooled down to room temperature with 2.2K/min cooling rate while maintaining the same pressure of 10$^{-6}$ mbar. 

The V$_2$O$_3$ thin film was characterized using X-ray diffraction (XRD), X-ray reflectivity (XRR), and Raman spectroscopy. XRD $\theta$-2$\theta$ and phi-scan measurements were performed using PAN analytical X'PERT high-resolution X-ray diffraction system equipped with a Cu anode. Raman measurements were performed using a micro  Raman spectrometer (Horiba Jobin-Yvon, France). 

The temperature-dependent resistivity and voltage (V) - current (I) curves of the V$_2$O$_3$ thin film are measured in the linear four-point probe configuration, using Keithley 2425 source meter and Keithley 2002 multimeter. The sample temperature was controlled using a Lakeshore 335 temperature controller. Temperature-dependent resistivity measurements in cooling and heating cycles were performed in the temperature sweep mode with a 2K/min temperature sweep rate with an applied current of 0.1 mA. Isothermal V-I characteristics were studied in continuous increment/decrement of current and measuring the corresponding voltage difference between two probe points. These isothermal measurements were performed under two experimental protocols: zero electric field (current) cooled and electric field (current) cooled protocols. In zero current cooled protocol, the isothermal V-I curves are obtained after the sample is cooled from high temperature (say 250 K) to the temperature of measurement in the presence of zero applied current.  In the current cooled protocol, the sample is cooled down from 250 K in the presence of an applied current ( say 0.15 A ) to the temperature of measurement, and isotherm is obtained after reducing the current to zero and then varying it in the sequence 0 A    $\rightarrow$ 0.15 A $\rightarrow$ 0 A $\rightarrow$ - 0.15 A $\rightarrow$ 0 A.   These two protocols of measurements are akin to zero (magnetic)field cooled (ZFC) and finite (magnetic)field cooled (FCC) experimental protocols widely used in the field of magnetism \cite{sbroy1}. To check the effect of joule heating, the isothermal measurements were also performed in pulsed current mode. The results obtained are quite similar in both continuous current and pulsed current modes, and thus rule out any significant role of the joule heating.

\begin{figure}
\centering
\includegraphics[width=8cm]{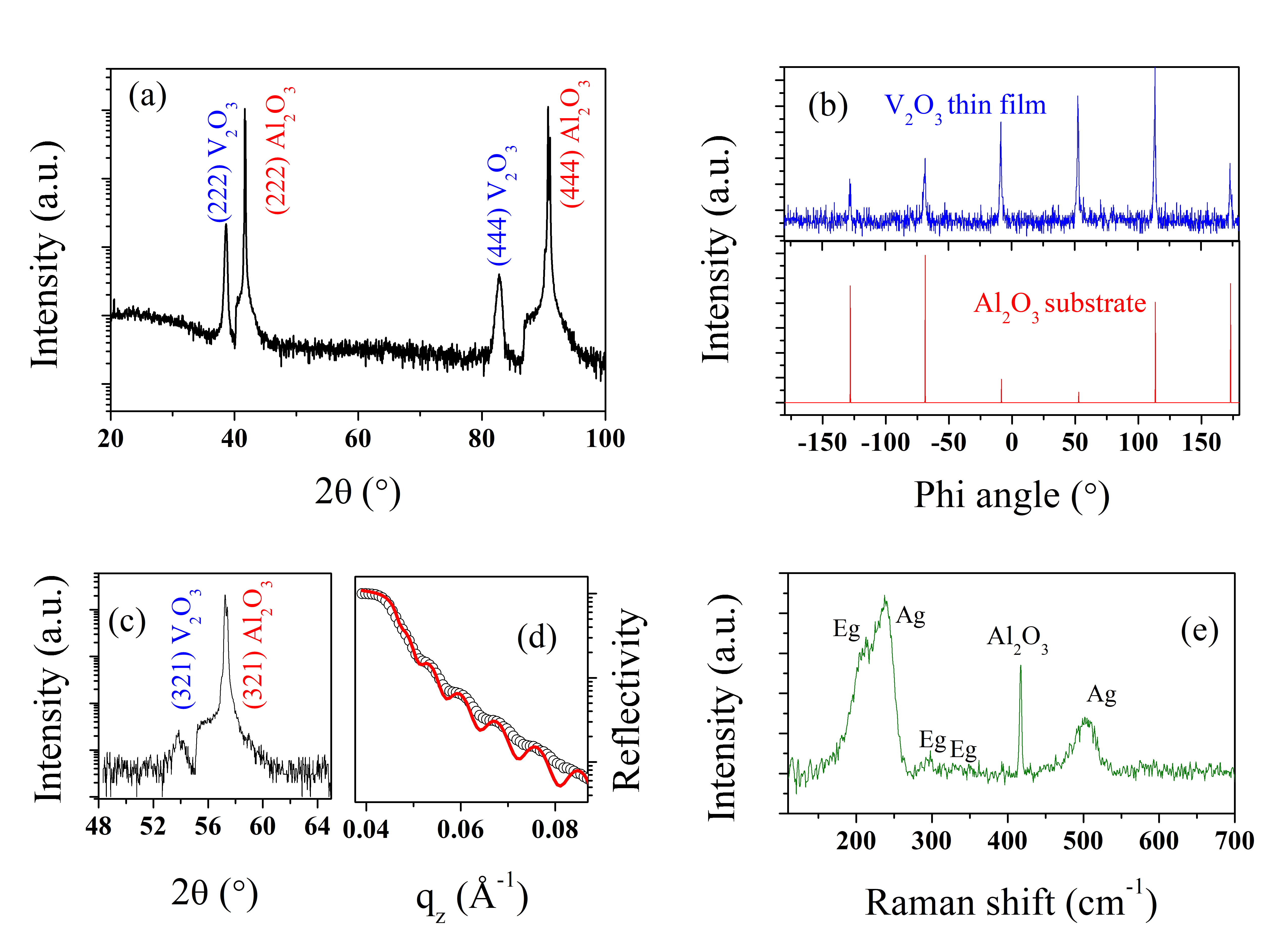}
\caption{Results of X-ray diffraction and Raman spectroscopy: (a) out-of-plane XRD $\theta$ - 2$\theta$ scan; (b) XRD phi scan on (321) reflection (phi scan on the Al$_2$O$_3$ substrate is also given at the bottom for comparison).  (c) in-plane XRD $\theta$ - 2$\theta$ scan around (321) reflection; (d) XRR results and (e) Raman spectroscopy results. }
\label{XRD}
\end{figure}  

\section{Results and discussions}

Fig.\ref{XRD}(a) presents out-of-plane XRD $\theta$-2$\theta$ scan, which reveals the (222) and (444) peaks of V$_2$O$_3$. This result of XRD measurement indicates that the film is grown along (222) direction of the rhombohedral R-3c structure on sapphire (222) substrate. The position of the (222) and (444) peaks matched with the bulk counterpart, indicating that the film is nearly relaxed. In-plane epitaxy is further confirmed using an asymmetric phi scan on (321) reflection. Fig.\ref{XRD}(b) shows the phi scan on V$_2$O$_3$ thin film along with the phi scan of Al$_2$O$_3$ substrate. The six-fold reflection of substrate and film at the same phi angle indicates the hetero-epitaxial nature of the film. Fig.\ref{XRD}(c) presents the in-plane XRD $\theta$-2$\theta$ scan around the (321) plane. We have estimated the thickness of the film with the help of XRR study. Uniform deposition of the film is indicated by distinct oscillations observed in XRR pattern (Fig.\ref{XRD}(d)).  We estimated film thickness to be $\approx$ 70 nm from the fitting of the XRR pattern. The microstructure of the deposited thin film V$_2$O$_3$ was further checked with a micro Raman spectroscopy study at room temperature (Fig.\ref{XRD}(e)).  From group theory, 2A$_g$+5E$_g$ are the only Raman active modes in the rhombohedral crystal structure of V$_2$O$_3$. Positions of the Ag (238, 502 cm$^{-1}$) and Eg modes (210, 294, 334 cm$^{-1}$) agreed well with the previous reports of Raman spectroscopy study in V$_2$O$_3$ \cite{tat,shvets}.  

\begin{figure}
\centering
\includegraphics[width=8cm]{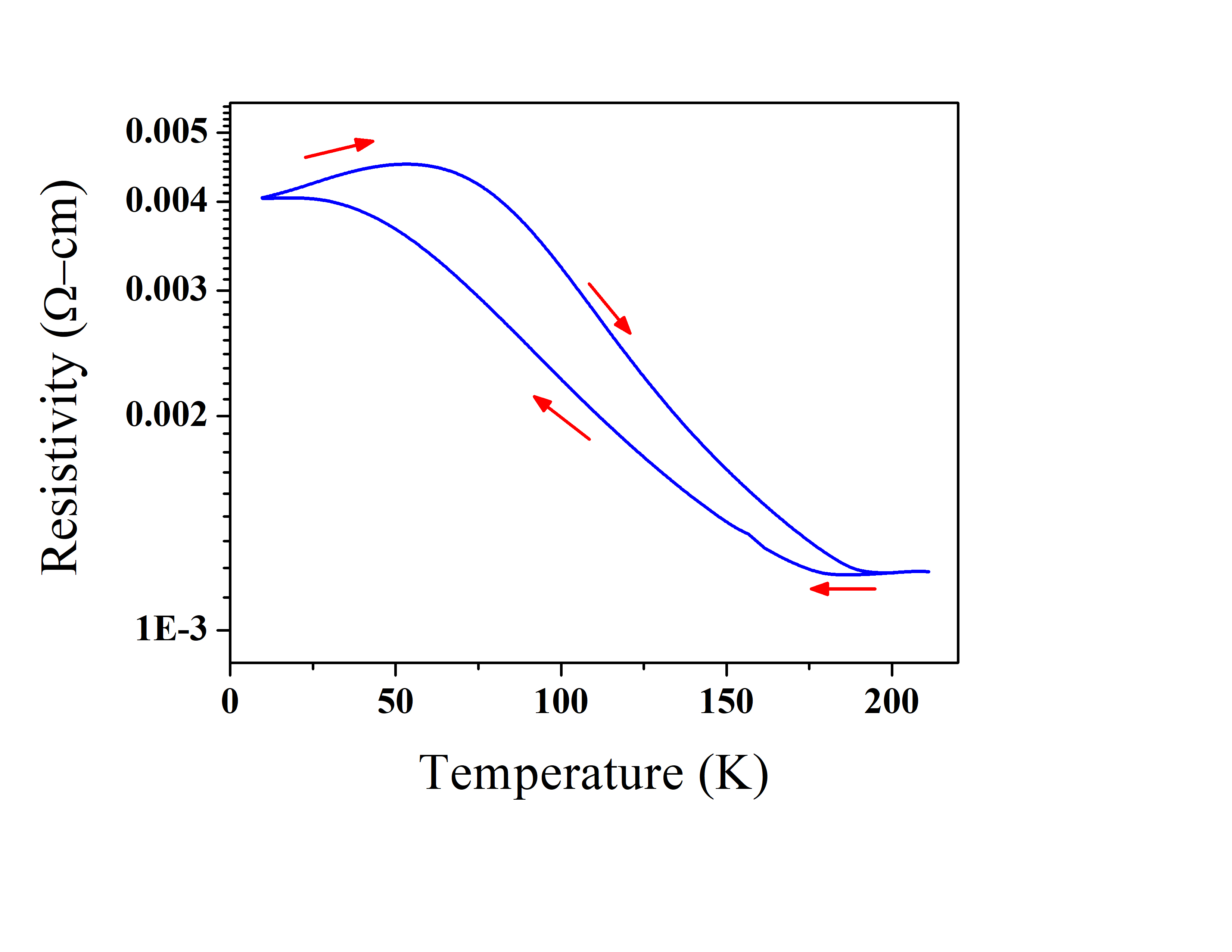}
\caption{Resistivity ($\rho$) versus temperature plot of V$_2$O$_3$ thin film obtained by variation of temperature both in the cooling and warming cycle. The driving current in the sample is 0.1 mA.}
\label{RvsT}
\end{figure}  

Fig. \ref{RvsT} presents temperature (T) dependence of resistivity ($\rho$) in V$_2$O$_3$ thin film between 210 K and 5 K. The driving current in the sample was 0.1 mA. With decreasing temperature, the sample undergoes a Mott metal-insulator transition around 180 K. This Mott transition temperature agrees well with the values reported in the literature both for bulk and thin film samples of V$_2$O$_3$ \cite{imada,macl}. In the warming-up cycle from the low temperature side, the Mott transition in our present sample is completed around 200 K, and thus the transition is marked with a distinct and broad thermal hysteresis. Such a broad transition is typically associated with a disordered influenced first-order phase transition, where instead of a single thermodynamic transition temperature a landscape of transition temperatures is introduced by quenched disorders in the system   \cite{imrywortis, sbroy1,sbroy2}. In an earlier study, a 160 nm V$_2$O$_3$ film grown on (001) sapphire showed a broadened Mott metal-insulator transition; the broadening was attributed to the intrinsic effect of self-straining \cite{kalch}. The same study also reported results on V$_2$O$_3$ films grown on various other orientations of sapphire substrates showing sharper Mott transition \cite{kalch}.  The present  epitaxial thin film of V$_2$O$_3$ grown on (222) sapphire is relatively relaxed, but the observed broadening of the in-plane as well as out-of-plane XRD $\theta$-2$\theta$ and $\phi$ scan indicates that structural disorders like line or plane dislocations may be present in the sample. Within the framework of Imry-Wortis model \cite{imrywortis}, such quenched disorders give rise to the broadening of a first-order phase transition. The magnitude of change in resistivity associated with Mott metal-insulator transition in our present sample is quite similar to that observed by Kalcheim et al in their  160 nm V$_2$O$_3$ film grown on (001) sapphire \cite{kalch}. It may also be noted here that the magnitude of the change in resistivity associated with the temperature-induced Mott metal-insulator transition in doped V$_2$O$_3$ samples is considerably smaller in comparison to the pure V$_2$O$_3$ \cite{mcw1}.  In the low-temperature region there is finite drop in the resistivity in the present V$_2$O$_3$ film in the warming-up cycle (see Fig. 2). Such anomalous resistivity drop in a much more pronounced form has been reported earlier in V$_2$O$_3$ film grown on (001) sapphire and tentatively attributed to a suspected reentrant metallic phase \cite{kalch}.  There is a structural change associated with the Mott metal-insulator transition in V$_2$O$_3$ \cite{imada}. Raman spectroscopy study at various temperatures below the Mott transition temperature confirmed the same in our present thin film V$_2$O$_3$ sample ( see Fig. S1 in \cite{binoy} ).

\begin{figure}
\centering
\includegraphics[width=8cm]{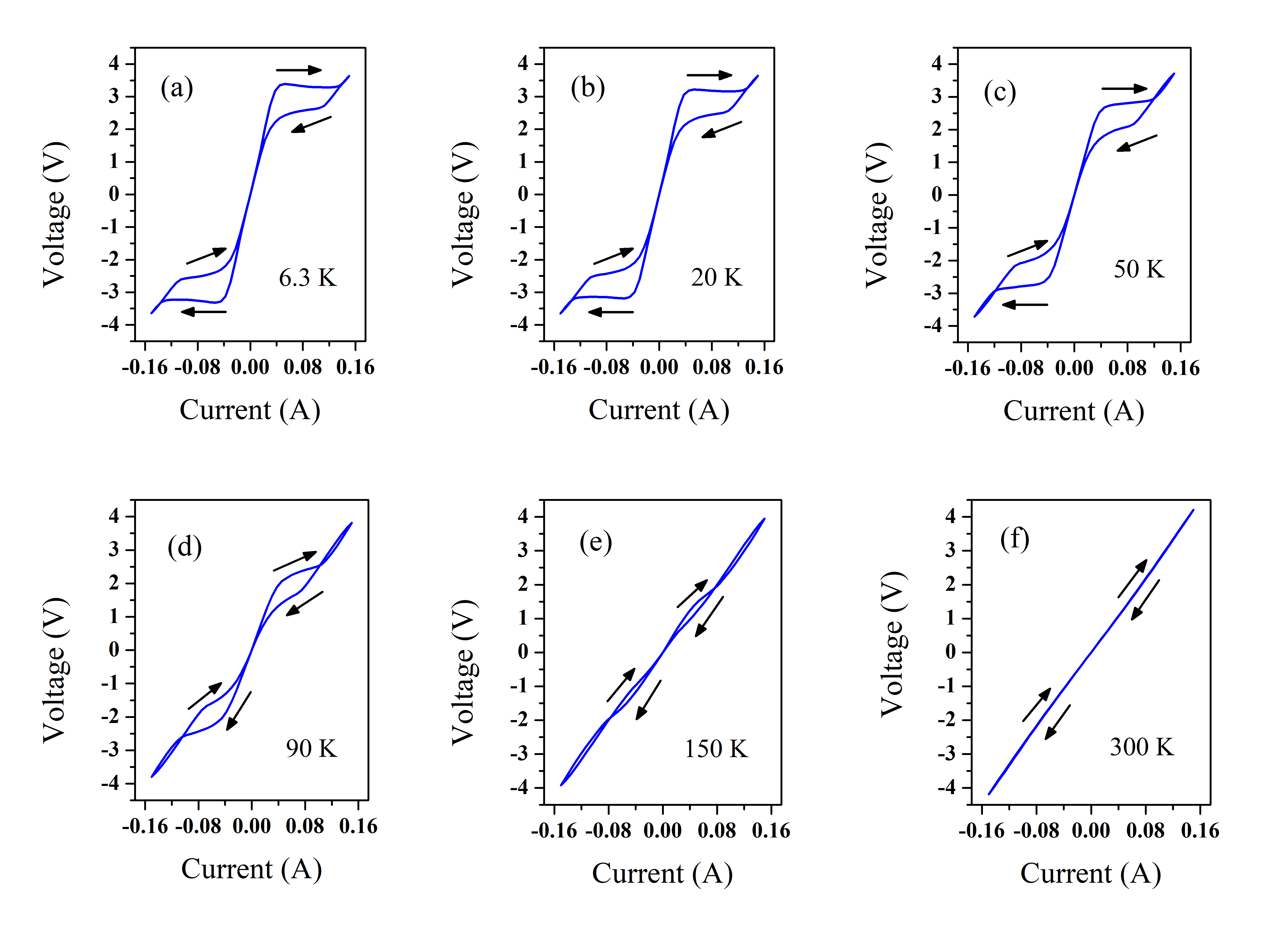}
\caption{Isothermal voltage (V) versus current (I) plots for V$_2$O$_3$ at various temperatures.}
\label{IVC1}
\end{figure}

The electric field-driven formation of a conductive state in a conventional band insulator system is based on the well-known Landau-Zener mechanism of quantum tunneling across the insulating band gap \cite{zener}. The rigid band gap is determined by the chemical composition and lattice structure of the materials, which needs to be overcome by the applied electric field. This sets a lower bound of the threshold electric field and limits the density of excited carriers promoted across the gap. Mott insulators with the collective nature of their gap can be ideal candidates to overcome this limit and provide a potential alternative to band semiconductor-based microelectronics. In recent times electrically induced insulator-to-metal transition in various Mott insulators has drawn much interest \cite{mazza,valm,kolch}.  To investigate the possibility of electrically driven Mott insulator to metal transition in the present V$_2$O$_3$ sample, in  Fig \ref{IVC1} we present isothermal V-I curves obtained at various temperatures both above and below the temperature region of the onset of Mott transition. The V-I curve at T = 300 K shows the typical linear metallic behaviour. This behaviour starts changing below the Mott transition temperature with the appearance of distinct non-linearity in the V-I characteristic accompanied by a hysteresis. A characteristic current or critical current can be estimated with considerable accuracy from the deviation from linearity in V-I curve (see Fig. S3 in \cite{binoy}). This current value increases with the decrease in temperature, and so does the non-linearity and the associated hysteresis.  As mentioned above, we have repeated some of these measurements by applying the current in pulsed mode, and we observed qualitatively the same behaviour. Thus the observed non-linearity and the associated irreversibility are robust and reproducible in different experimental cycles. Below the threshold current, the V-I curves are linear and reversible at all temperatures. Such pinched hysteresis loops in V-I curves are typically associated with resistive-switching memories or memristors \cite{chua,chua2}. Thus the observed non-linearity and hysteresis in the present V$_2$O$_3$ epitaxial thin-film can be utilized to build a volatile electrical resistive switching device. It may, however, be mentioned here that circuits with nonlinear inductors and capacitors can also give rise to such pinched hysteresis loops, which are not necessarily memristive systems \cite{biolek,jwu}.

\begin{figure}
\centering
\includegraphics[width=8cm]{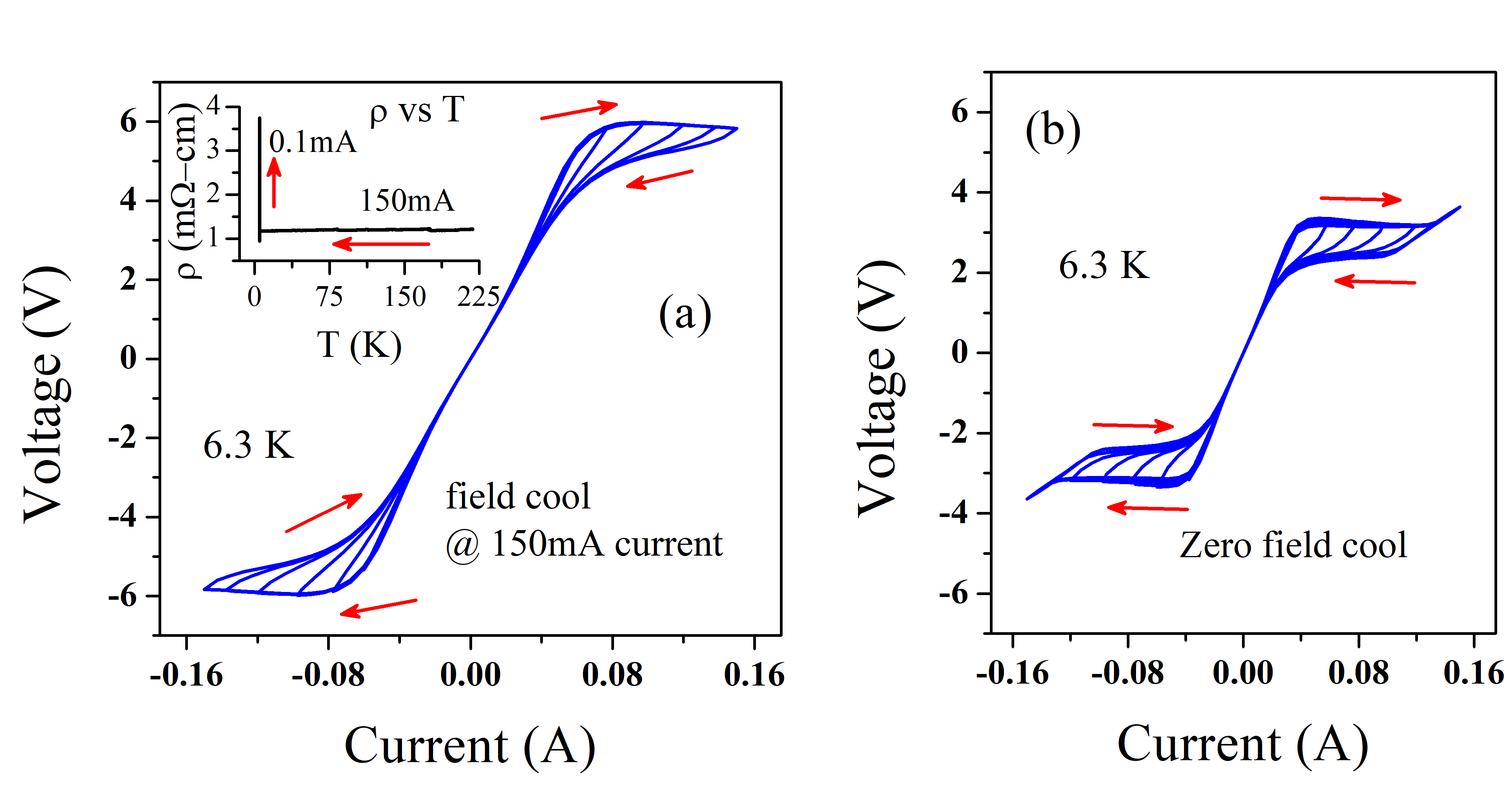}
\caption{Isothermal voltage (V) versus current (I) plots for V$_2$O$_3$ thin film along with minor hysteresis loops at T = 6.3 K obtained under (a) current cooled and (b) zero current cooled protocols. Inset of Fig. 4(a) shows the resistivity ($\rho$) versus temperature (T) plot obtained while cooling the sample from high temperature across the Mott transition with a measuring current of 0.15 A. After reaching the temperature 6.3 K, the resistivity was measured by changing the driving current in the sample to 0.1 mA while keeping the temperature constant. }
\label{IVC2}
\end{figure}  

There is much debate in the scientific community on the two competing mechanisms namely a purely electronic mechanism and an electro-thermal mechanism to explain the electric field-induced insulator to metal phase transition in Mott insulators \cite{mazza,valm,kolch,valle}. In systems like VO$_2$ and V$_2$O$_3$ where Mott transition takes place as a function of temperature, Joule heating due to current flow in the sample is an obvious possibility for inducing the transition. However, it has also been argued quite convincingly that the applied electric field can induce the insulator-metal transition non-thermally, without heating the material to its Mott transition temperature \cite{valm,stol1,stol2,dien,janod}

 Mazza et al \cite{mazza} with the help of dynamical mean-field theory argued that the electric breakdown of the Mott insulator occurred via a Mott insulator to metal first order phase transition (FOPT) characterized by an abrupt gap collapse. They showed that within the insulator-metal coexistence region, an electric field could drive a discontinuous transition from the insulator to a gap-collapsed metal at the threshold electric fields much smaller than those expected in a Zener breakdown. In the region away from insulator-metal coexistence, the electric breakdown occurs through a more conventional quantum tunneling across the Hubbard bands tilted by an electric field of significant strength. This work established a framework to understand the metallization of a Mott insulator, where the electric field drives a FOPT from the correlated Mott insulator to a gap-collapsed metal phase, preexisting as a metastable state at zero fields. del Valle et al \cite{valle} argued that the Mott insulator to metal transition in V$_2$O$_3$ could be triggered directly by the electric field, possibly by carrier injection into the conduction band and destabilization of the insulating phase. Joule heating plays a role in the growth and percolation of the metallic phase, but it is not the primary cause of metallization \cite{valle}. In our present experiment, we have ruled out the primary role of Joule heating by repeating some experiments in pulsed current mode. Furthermore, we have separately checked the increase in the temperature of the sample on application of a current of 0.15 A by putting a temperature sensor very near to the sample after stabilizing the temperature at 6.3 K. The observed rise in the temperature was 6.4 K. 

Defects can play an important role in the electric field-induced Mott transition as they may create in-gap trap states that release carriers by thermal activation \cite{kolch}. In this direction, Zener-tunneling is a well-studied mechanism for electric field-induced doping and destabilization of Mott insulators \cite{oka,oka2,yama,lee}. The required electric field within this mechanism is relatively high to enable the tunneling of carriers across the Mott gap, which induces a doping-driven transition. In our present work as well as the experimental studies on V$_2$O$_3$ by Kalcheim et al \cite{kolch},  the switching fields are quite
small so that direct Zener-tunneling can be ruled out. A similar effect, however, is possible by electric field-assisted activation of carriers from trap-states \cite{kolch}. In place of tunneling through the gap, charge trapped in defect in-gap states can be activated, which enables enough carriers to dope the Mott insulator and cause the insulator to metal transition. There exist theoretical studies, which indicate that defects and impurities can indeed reduce the switching field considerably than the one expected from standard Zener tunneling \cite{lee,sugi}. For a more detailed discussion on the important role of defects in the non-thermal microscopic mechanism of electric field-induced Mott transition in V$_2$O$_3$ the readers referred to the work of Kalcheim et al \cite{kolch} and the references therein. 

McLeod et al \cite{macl} experimentally established such spontaneously nanotextured coexistence of metal and correlated Mott insulator phases near the insulator to metal transition (around 160-180 K) associated with percolation in V$_2$O$_3$. It may be noted that such phase-coexistence is a universal feature of a disorder-influenced FOPT in general, and has been studied extensively in various classes of magnetic materials \cite{sbroy1}. Such a temperature and magnetic field-driven disorder-influenced FOPT with its associated features (phase coexistence, thermo-magnetic irreversibility, etc.) can give rise to various kinds of functionalities namely giant magneto-caloric effect, magneto-striction, magnetic shape memory effect, etc \cite{sbroy1}. In the same line of approach, here we use an electric field (current) here instead of the magnetic field. In Fig. \ref{IVC2} we present V-I curves along with minor hysteresis loops (MHLs) at T = 6.3 K in the present V$_2$O$_3$ sample, obtained in an electric field (current)-cooled mode and also in the conventional zero-electric field (current)-cooled mode. The MHLs are drawn by limiting the isothermally applied current to various values less than the maximum applied current of 0.15 A. There is a distinct difference between the V-I curves obtained in current-cooled and zero-current-cooled modes. In the current-cooled mode, the high-temperature metallic state persists as a supercooled metastable state to a temperature region well below the onset temperature of the first-order Mott metal-insulator transition. This is exemplified in the inset of Fig. 4(a), which shows the resistivity of the present V$_2$O$_3$ sample as a function of temperature with a measuring current of 0.15 A, obtained by cooling the sample from high temperature across the Mott transition. There is no Mott transition and metallic behaviour is retained down to 6.3 K.   The prominent existence of this high-temperature state with rhombohedral structure in the current cooled mode down to 6.3 K, is also confirmed by the concomitant Raman spectroscopy study \cite{binoy}. However, this non-equilibrium metallic state is quite fragile at low temperature. It will be susceptible to any fluctuations introduced during the process of measurement and will tend to get transformed to the equilibrium insulating state. This can be observed in the inset of Fig. 4(a). After reaching 6.3 K the driving current in the sample is changed from 0.15 A to 0.01 mA (which was the value of current used to measure the resistvity presented in Fig. 2) while keeping the temperature constant. The resistivity rises rapidly indicating transformation of the non-equilibrium metallic state towards a stable insulating state.  This process breaks some percolative conducting paths and gives rise to different electrical responses in the system with a relatively high resistivity (see Fig. 4(b)).  We have observed qualitatively the same kind of V-I curves with minor hysteresis loops at T = 30 and 50 K (shown in the Fig. S2 of supplementary material \cite{binoy}), but the results are not included in the main text for the sake of conciseness \cite{binoy}. This electro-thermal history effect is quite akin to thermo-magnetic history effects observed between zero magnetic-field cooled (ZFC) and magnetic-field cooled (FC) states in various magnetic systems showing first-order phase transition \cite{sharma,chatt,sbroy1}. The magnetic response in many such systems, however, is a cumulative effect, and it is straightforward to understand the experimentally observed quantities in terms of the supercooling of the high-temperature magnetic state across the phase transition point and its persistence as a fragile metastable state. In the present case of V$_2$O$_3$, the coexistence of the transformed Mott-insulating state and the supercooled high-temperature metastable metallic state gives rise to a complex insulator-metal network configuration, which is expected to be different from the one obtained after applying the electric field in the zero current cooled state.  The MHLs and associated magnetic memory effect have also been extensively studied in various magnetic systems, and they provide macroscopic evidence of the phase coexistence in a disorder broadened first order phase transition \cite{man1,man2,sbroy1}. Within a similar framework, the electric-field induced MHLs in the present case of V$_2$O$_3$, can be suitably utilized for memory applications with a relatively low power ( $<$ 1 W) requirement. 

We will now use a resistor network model \cite{stol1,stol2} to explain the switching feature in the V-I curves observed in our present experimental study. The cells in this network are modeled by four resistors with value $\it{R_{MI}}$ when it is in the Mott insulator (MI) state or $\it{R_{CM}}$ when it is in the correlated metal (CM) state, with  $\it{R_{CM}}<$ $\it{R_{MI}}$. Fig. 5(a) presents the array of such cells connected to a current source $\it I$. A small portion of the sample (of at least a few tenths of nanometers) is represented by each cell.   In the low applied electric field (current) V$_2$O$_3$ is in the stable-MI state having lower energy and that is defined as the reference energy E$_{MI} \equiv$ 0. An energy barrier $\it{E_B}$ separates the MI and the CM state in each cell. A small portion of the sample (of at least a few tenths of nanometers) is represented by each cell, such that the local resistivity may have a well-defined value \cite{stol1,stol2}. In the critical region of field-induced transition, a local electric field $\epsilon$ destabilizes the electronic configuration of a cell in the MI state by effectively reducing the $\it{E_B}$ barrier. This results in a probability for the local conversion of the MI state toward the CM state \cite{stol2}:
\begin{equation}
P_{CM}(x,y) = exp\left(-\frac{E_B - q\vert \epsilon(x,y)\vert}{k_BT(x,y)}\right)
\end{equation}

When a cell is converted into the highly conductive CM state with E$_{CM}$, its internal electric field becomes negligible, and the cell will then have a probability \cite{stol2}:
\begin{equation}
P_{MI}(x,y) = exp\left(-\frac{E_B - E_{CM}}{k_BT(x,y)}\right)
\end{equation}
for switching back to the MI state.

At the onset of the critical regime of electric field-induced Mott transition, the system consists of largely a homogeneous MI state and a few isolated CM cells. With the increase in the applied electric field, a first-order phase transition from the MI to CM state takes place. In this higher current regime as a part of this first-order transition process, at a threshold current, a high local field region will be the nucleation point for the sudden formation of a percolative metallic cluster that connects the electrodes. This electronic model provides a reasonable explanation of the experimentally observed resistive switching in our present epitaxial thin film V$_2$O$_3$ sample above a threshold current.

\begin{figure}[htb]
\begin{center}
\includegraphics[width=0.6\textwidth]{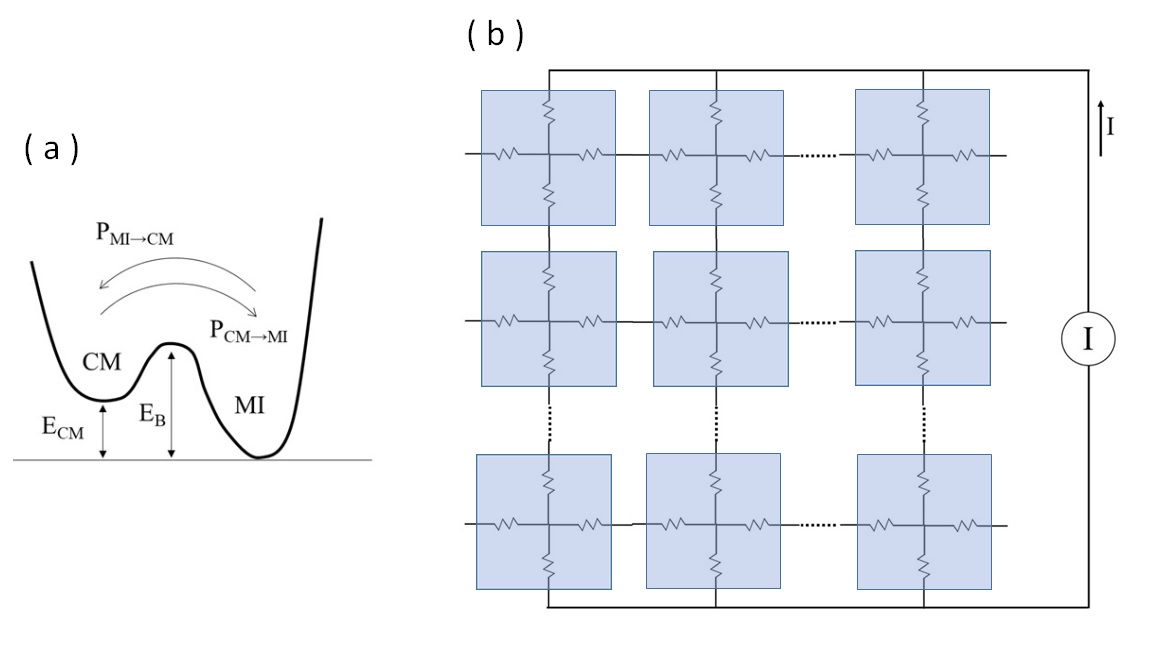}
\end{center}
\caption{Schematic representation of a pure electronic model of the resistor network along with the energy barrier between the Mott insulator state and the correlated metal state (Reproduced from Ref. \cite{stol2} with the permission of American Physical Society).}
\end{figure}

We have not observed any global change in temperature due to the applied critical current in our present V$_2$O$_3$ thin film sample to be sufficient enough to drive a temperature-induced MI to CM transition, but at the same time it remains a fact that the resistivity of the Mott insulators is quite sensitive to local changes in temperature. So, in any realistic system, a combination of the electronic and the thermal models is necessary to capture the interplay between the Mott insulator-metal transition and the local Joule heating. To this end the switching probabilities $\it{P_{CM}}$ and $\it{P_{MI}}$ (see eqns. 1 and 2) depend on the local temperature. However, within this combined model it is observed that the increase in temperature is minimal before the electric field-induced transition \cite{stol2}. The switching behavior is dominated by the electric-field-driven MI-CM transition. Post switching, first there may be a considerable increase in the temperature of the filament, and that is followed by the stabilization of a state with a higher conductance after a current dependent time-scale. Thus, both electronic and thermal mechanisms play their role but at different time scales, \cite{stol2}. The current regulation after switching is observed at short time scale, while at longer times this stabilization may be lost due to filament Joule heating.

\begin{figure}
\begin{center}
\includegraphics[width=0.6\textwidth]{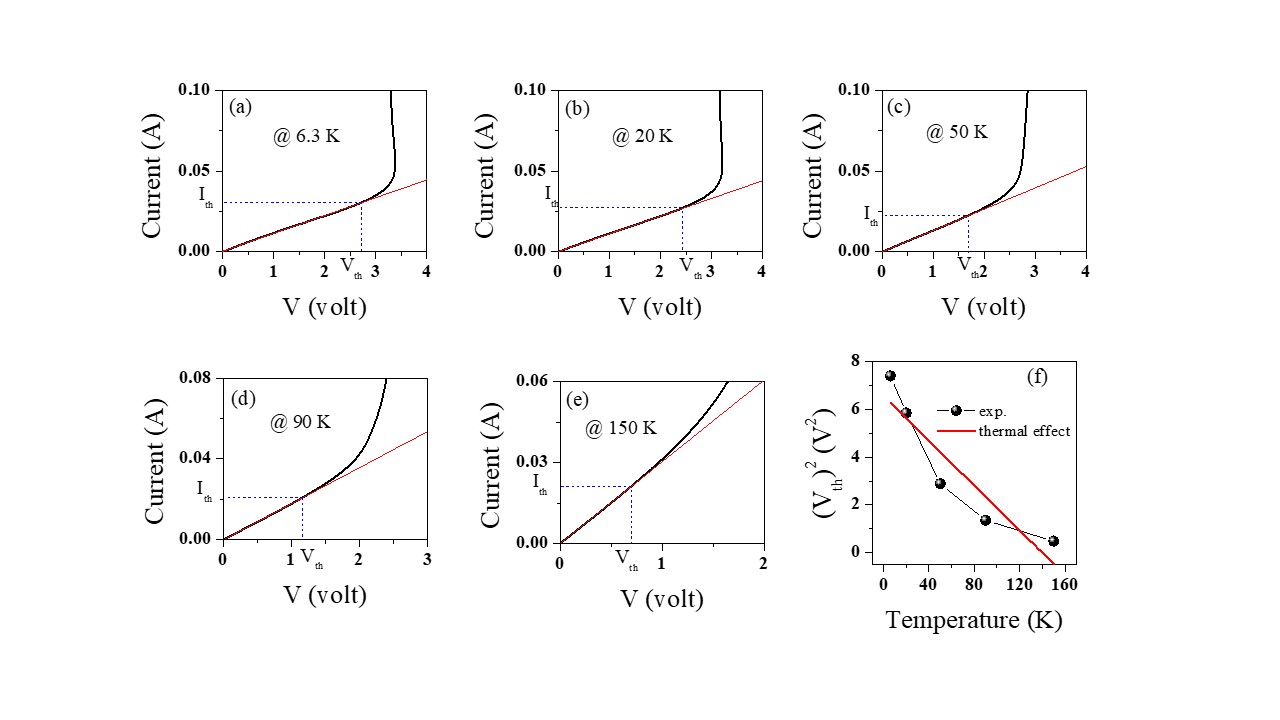}
\end{center}
\caption{(a-e) Isothermal V-I characteristics at T = 6.3 20, 50, 90 and 150 K obtained under zero current cooled protocol, highlighting the deviation from the linear behaviour in V-I curve. (f) Variation of V$_{th}$ with temperature. Red line represents the linear fit with V$_{th}^2$ = KR$_{ins}$(T$_{MIT}$-T) corresponding to the thermally induced resistive collapse (see main body text for details).}
\end{figure}

More recently Rocco et al \cite{rocco}  have studied theoretically how a Mott insulator subject to a strong applied voltage can undergo an inhomogeneous insulator-to-metal transition with formation of metallic filaments within the insulating bulk. It is shown that resistivity collapse due to an external electric pulse is a time dependent phenomenon and the incubation time depends on the pulse strength; stronger pulse produces first switching (deterministic switching). In the present experimental study we have observed a distinct non-linearity in the V-I characteristic accompanied by a hysteresis in the temperature region below the onset of Mott metal-insulator transition (see Fig. 3).  A characteristic voltage can be assigned and estimated fairly accurately from the deviation from linearity in the V-I curve obtained at various temperatures (see Fig. 6 (a)-(e)). The required lowest voltage to observe a finite incubation time (here it is 1sec) is denoted by the threshold voltage, V$_{th}$.  We have done all the measurements isothermally under a constant temperature, and we can say that the observed temperature variation of the threshold voltage is the temperature variation of the minimum required voltage for resistivity collapse. The temperature evolution of the V$_{th}$ is shown in Fig. 6(f). Theoretically estimated thermally induced V$_{th}$ follows the relation V$_{th}^2$= KR$_{ins}$(T$_{MIT}$-T), where K and R$_{ins}$ are the thermal conductivity and resistivity in insulator phase \cite{rocco}. There is a clear deviation of the experimental obtained V$_{th}^2$ from the linear behavior (red line) corresponding to the thermally induced effect.This further reinforces the idea that the resistive collapse arises mainly due to a non-thermal electronic effect, but at the same time the thermal effect does play a role by assisting the electronic effect in the process of resistive switching.

\section{Summary and conclusion}
In summary, we have observed a primarily electric field-induced Mott-insulator to correlated  metal transition in epitaxial thin film of Mott insulator V$_2$O$_3$ deposited on (222) Al$_2$O$_3$ substrate in the temperature regime below the Mott transition. This first-order Mott transition is accompanied by interesting electro-thermal history effects, which have been investigated by following different measurement paths in the electric field (E) - temperature (T) phase space. The results are distinctly different depending on whether the sample is cooled to the temperature of measurement in the presence or absence of an applied electric field (current). A generalized framework of disorder-influenced first order phase transition \cite{imrywortis} in combination with a resistor network model \cite{stol1,stol2} have been used to rationalize the observed experimental features. The characteristic electro-thermal-electric history effects result in volatile resistive switching and distinct memristive behaviour in V$_2$O$_3$. Thus it is a promising candidate for novel technologies like neuromorphic computing \cite{pickett,zhou,kumar} and optoelectronics \cite{liu, but} .

%\section{Supplementary materials}
The supplementary materials provides additional V-I curves obtained in current cooled mode and Raman spectroscopy results obtained both in zero current cooled and current cooled mode at various temperatures to support the main results of the paper. 

%\{Acknowledgement}
Binoy De and S. B. Roy acknowledge financial support obtained from the Raja Ramanna Fellowship program sponsored by DAE, Government of India.

%\{Conflict of interest}
The authors have no conflicts to disclose.

\end{document}